\documentclass[12pt,preprint2]{aastex}
\usepackage{emulateapj5,apjfonts,onecolfloat5}

\shorttitle{Discovery of a New Magnetar Candidate}
\shortauthors{Ibrahim et al. 2003}

\begin{document}
\twocolumn[

\doublespace

\title{Discovery of a New Transient Magnetar Candidate: XTE~J1810--197}

\author{A.~I.~Ibrahim,\altaffilmark{1,2} C.~B.~Markwardt,\altaffilmark{1,3} J.~H.~Swank,\altaffilmark{1} 
S.~Ransom,\altaffilmark{4,5} M.~Roberts,\altaffilmark{4,5} V.~Kaspi,\altaffilmark{4,5} 
P.~M.~Woods,\altaffilmark{6} S.~Safi-Harb,\altaffilmark{7} S.~Balman,\altaffilmark{8} W.~C.~Parke,\altaffilmark{2} 
C.~Kouveliotou,\altaffilmark{9} K.~Hurley,\altaffilmark{10} T.~Cline\altaffilmark{1}}

\affil{
$^{1}$NASA Goddard Space Flight Center, Laboratory for High Energy Astrophysics, Greenbelt, MD 20771 \\ 
$^{2}$Department of Physics, The George Washington University, Washington, D.C. 20052 \\
$^{3}$Department of Astronomy, University of Maryland, College Park, MD 20742 \\
$^{4}$Department of Physics, Rutherford Physics Building, McGill University, Montreal, QC H3A 2T8, Canada \\
$^{5}$Center for Space Research, Massachusetts Institute of Technology, Cambridge, MA 02139 \\
$^{6}$Universities Space Research Association, NSSTC, SD-50, 320 Sparkman Drive, Huntsville, AL 35805 \\ 
$^{7}$NSERC UFA fellow, Department of Physics \& Astronomy, University of Manitoba, Winnipeg, MB R3T 2N2, Canada \\
$^{8}$Department of Physics, Middle East Technical University, Ankara, Turkey \\
$^{9}$NASA/MSFC, NSSTC, SD-50, 320 Sparkman Drive, Huntsville, AL 35805 \\
$^{10}$Space Sciences Laboratory, University of California at Berkeley, Berkeley, CA 94720-7450 \\
\vspace{.2cm}
}

\begin{abstract}

We report the discovery of a new X-ray pulsar, XTE~J1810--197. The
source was serendipitously discovered on 2003 July 15 by the {\em Rossi X-ray Timing
Explorer (RXTE)} while observing the soft gamma repeater SGR~1806--20. 
The pulsar has a 5.54 s spin-period and a soft spectrum (photon index $\approx 4$).
We detect the source in earlier RXTE observations back to 2003 January.
These show that a transient outburst began between 2002 November 17 and 2003 January 23
and that the pulsar has been spinning down since then, with a high 
rate $\dot{P} \approx 10^{-11}$ s s$^{-1}$ showing significant timing noise, but no evidence 
for Doppler shifts due to a binary companion.
The rapid spin-down rate and slow spin-period
imply a super-critical magnetic field $B=3 \times10^{14}$ G and a
young characteristic age $\tau \leq 7600$ yr. 
These properties are strikingly similar to those of anomalous X-ray pulsars and soft gamma
repeaters, making the source a likely new magnetar. 
A follow-up $Chandra$ observation provided a $2\farcs5$ radius error circle
within which the 1.5 m {\em Russian-Turkish} Optical Telescope {\em RTT150}  found
a limiting magnitude of $R_c=21.5$, in accord with other recently reported limits. 
The source is present in archival {\em ASCA} and {\em ROSAT} data
as well, at a level 100 times fainter than the $\approx$ 3 mCrab seen in
2003. This suggests that other X-ray sources that are currently in a state 
similar to the inactive phase of XTE~J1810--197 may also be unidentified magnetars 
awaiting detection via a similar activity.

\end{abstract}

\keywords{Pulsar: Individual (XTE~J1810--197) --- Stars: Magnetic Fields --- Stars:
Neutron --- Stars: Magnetar --- X-Rays: Bursts}

]

\section{Introduction}

Several hundred X-ray pulsars have been discovered to date.
Some are powered by their own rotational energy or residual surface heat and others 
by accretion. The two subgroups of
anomalous X-ray pulsars (AXPs) and soft gamma repeaters (SGRs)
are remarkably distinct from the rest and similar to each other.  
They rotate relatively slowly with
spin periods in the narrow range $P \sim 5-12$ s and spin-down rather
rapidly at $\dot P \sim 10^{-11}$ s s$^{-1}$. 
Both are radio-quiet, sources of
persistent X-ray emission ($L \sim 10^{34}-10^{36}$ erg s$^{-1}$) and
short ($<0.1$ s), bright ($L_{peak} > L_{EDD}$) bursts of X-rays
and soft $\gamma-$rays. They are peculiar in that there is no evidence 
of a binary companion or a remnant accretion disk to power their emission,
although it is several orders of magnitudes higher than can be provided by
their rotational energy.
Nine sources are currently firmly identified, including four SGRs and 
and five AXPs (See Hurley 2000 and Mereghetti et al. 2002). Four candidates
need confirmation.

The magnetar model provides a coherent picture for SGRs and AXPs, in
which their radiation is powered by a decaying super-critical magnetic
field, in excess of the quantum critical field $B_c =
4.4\times10^{13}$ G (Duncan \& Thompson 1992; Thompson \& Duncan 1995).  
Evidence for magnetars has come from the long spin-period and high spin-down rate (Kouveliotou et
al. 1998; 1999; Vasisht \& Gotthelf 1997), the energetic burst emission (Paczynski 1992; 
Hurley et al. 1999; Ibrahim et al 2001), and the lack of
binary companion or accretion disks (Kaplan et al. 2001). 
Further evidence for magnetar strength fields has recently come from 
spectral line features (Ibrahim et al. 2002; Ibrahim, Swank \& Parke 2003). For one case a
pulsed optical counterpart appears consistent with
being the neutron star itself (Hulleman et al. 2000; Kern \& Martin 2002). 
Until recently only SGRs were observed
to burst. The recent bursting activity from two AXPs unified the two
families of objects in the magnetar framework (Gavriil, Kaspi \& Woods 2002;
Kaspi et al. 2003). Alternative models such as fossil accretion
(Chatterjee et al. 2000; Marsden et al. 2001) and strange quark stars
(Zhang et al. 2000) do not appear to explain all observational evidence as well. 

Here we present the discovery of a new X-ray pulsar whose properties, in outburst, are
consistent with those of AXPs and SGRs.  We discuss the implications
of this finding for our understanding of the characteristics and
population of magnetars.

\section{Observations and Results}

\subsection{A New X-ray Pulsar near SGR 1806-20}

Following the {\it Interplanetary Network (IPN)} report of renewed burst activity from 
SGR~1806--20 on 2003 July 14 (Hurley et al. 2003), we observed the source on
July 15 with the Proportional Counter Array (PCA) onboard RXTE. 
PCA data in the event-mode configuration E\_125US\_64M\_0\_1S 
were collected from the operating PCUs (0, 2 \& 3), corrected to the solar system barycenter,
selected to be in the 2-8 keV range, and binned in 0.125 s intervals.
PCA detector layers are not differentiated in that mode.
A strong periodic signal with a barycentric period of 5.540(2) s 
was clearly identified in the first 2.6 ks of the data, with a chance
probability of $2.5\times10^{-12}$ (Ibrahim et al. 2003; see Fig.
1). The large discrepancy between this unexpected pulse period and the
known 7.5 s pulse period of SGR 1806-20 implied the presence of a new
X-ray pulsar in the PCA $1^\circ.2$ field of view.

\begin{center}
\vspace{-0.2cm}
\includegraphics[scale=0.3]{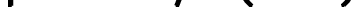}
\figurenum{1}
\vspace{-0.3cm}
\figcaption{Fast Fourier Transform power spectrum of the RXTE PCA July 15 
observation of the field of SGR 1806-20 showing a highly 
significant periodic signal at $0.18052(6)$ Hz (note that the $\approx 0.13$ Hz 
signal due to SGR 1806-20 is not detected here). 
The inset shows the epoch folded pulse profile in 10 phase bins. 
Errors quoted correspond to the $3\sigma$ confidence level.
}
\end{center}

\subsection{Source Position and Optical Counterpart}

A PCA scanning observation was performed on July 18, following a path
that covered a region surrounding SGR 1806--20. During scans, the
count rates due to individual sources are modulated by the response of
the PCA collimator. The resulting light curves are subtracted of their
internal background (using the ``CM'' L7 background model), and are
fitted to a model of known and unknown sources, convolved with the
collimator response. For unknown sources, a trial position is assumed
and adjusted until the best fit is achieved.  The sources included in
this fit were the new source, SGR 1806--20, the galactic ridge, and an
overall diffuse level.  The spatial distribution of the galactic ridge
emission in the field of view is not precisely known; an unresolved ridge
at $0^\circ$ latitude was assumed.  The best fit position and $3\sigma$
contour obtained for the position of the new source, designated
XTE~J1810--197, are shown in Fig. 2 (Markwardt, Ibrahim \& Swank
2003).

A follow up $Chandra$ observation with the High Resolution Camera
(HRC) on August 27 localized the source precisely to $\alpha=18^{\rm
h}09^{\rm m}51^{\rm s}.13$ and $\delta=-19\degr 43\arcmin 51\farcs7$
(J2000), with an error circle radius of 2\farcs5 (Gotthelf et al. 2003a;
2003b). Pulsations of the HRC data definitively identified the source.
The HRC position is consistent with the $ROSAT$ and $ASCA$ sources 
1RXS 180951.5--194345 \& AX J180951--1943 (Bamba et al. 2003).

The HRC position is 14\arcmin\ from the best fit PCA position.
Typically, accuracies of 1--2$\arcmin$ have been obtained in past
scans.  The presence of the diffuse galactic ridge and other faint
sources in the field of view --- in particular G11.2--0.3 --- resulted
in large systematic errors, for which a priori estimates were
difficult.

We observed the $Chandra$ HRC error box of XTE~J1810--197 with the 1.5
m $Russian-Turkish$ Telescope, $RTT150$ (Antalya, Turkey) on 2003
September 3 and 6.  Optical Cousins R filter images of the field
around the source were obtained using the ANDOR CCD ($2048\times2048$
pixels, $0.24\arcsec$ pixel scale and $8\arcmin\times 8\arcmin$ Filed
of View) with 15 min exposure times (3 frames).  Seeing was about
$2\arcsec$. 
We do not detect a counterpart to a limiting magnitude of
21.5 ($2\sigma$ level) in the $R_c$ band, comparable to
the limits in $V(22.5)$, $I(21.3)$, $J(18.9)$, and $K(17.5)$ obtained by Gotthelf et al. 
(2003b).

\begin{center}
\vspace{-0.2cm} \includegraphics[scale=0.3]{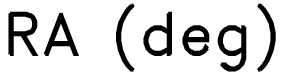}
\figurenum{2}
\vspace{-0.2cm}
\figcaption{The PCA field of view 
during the SGR 1806-20 pointing, 
showing the neighborhood of SGR 1806-20, including XTE~J1810--197, 
the supernova remnant G11.2-0.3 that 
contains the pulsar PSR~J1811--1925, and the potential SGR 1808-20 (Lamb et al. 2003).
The positions of XTE~J1810--197 from the PCA scan and HRC observations are indicated. 
Also shown is the 3$\sigma$ PCA error contour, with semi-major axes 
of 5.5\arcmin\ and 10\arcmin.
} 
\end{center}

\subsection{Long Term Light Curve: A Transient Source}

XTE~J1810--197 was consistent with a previously unidentified source
that had been present in the PCA monitoring program of the galactic
bulge region since 2003 February. In the program, a region of
approximately 250 square degrees around the galactic center has been
scanned by the PCA twice weekly since 1999 February, except for
several months per year, when sun and operation constraints
interfere (Swank \& Markwardt 2001). The scan pattern is a zig-zag which alternates semi-weekly
between primarily north-south and east-west.  XTE~J1810--197 is
covered in the north-south scans only.  At the end points of each scan
the PCA dwells for $\approx 150$ seconds, and XTE~J1810--197 is near
the center of the PCA field of view of one of these points.  As
discussed below, pulsations were observed during these brief points,
confirming the identification of the source.




Fig. 3 shows the 2002--2003 light curve of XTE~J1810--197, when fixed
at the $Chandra$ position.  Clearly XTE~J1810--197 became active
sometime between 2002 November and 2003 February.  The distribution of
1999--2002 pre-outburst fluxes allow us to place a $3\sigma$ upper
limit on previous outbursts of $<$ 2 ct/s/PCU or 1 mCrab (2--10 keV)
from the baseline level, as long as the outburst did not fall in an
observing gap (the maximum gap was 3 months).

The flux decay can be fitted to power-law or exponential models.  For
the exponential model, the e-folding time is $269 \pm 25$ days. The
power-law model has the potential of retrieving the epoch at which the
outburst began. Assuming the flux is proportional to $
((T-T_0)/(52700-T_0))^{-\beta}$, at time $T$ and with outburst time
$T_0$ in MJD, $\beta = 0.45-0.73$ were acceptable ($1 \sigma$), with
$52580 \le T_0 \le 52640$, that is, 2002 November 2 to 2003 January 1.
We have additional information from observations of the nearby source
PSR~J1811--1925 (Obsids 70091-01, 80091-01). An observation on 2002
November 17 (MJD 52595) showed that the pulsations were not detected,
while they were by 2003 January 23 (MJD 52662).  The {\em ROSAT} and
{\em ASCA} detections in 1993, 1996, and 1999 were consistent with
fluxes $\leq 0.02$ mCrab (Gotthelf et al. 2003b).


\begin{center}
\vspace{-0.2cm}
\includegraphics[scale=0.3,angle=90]{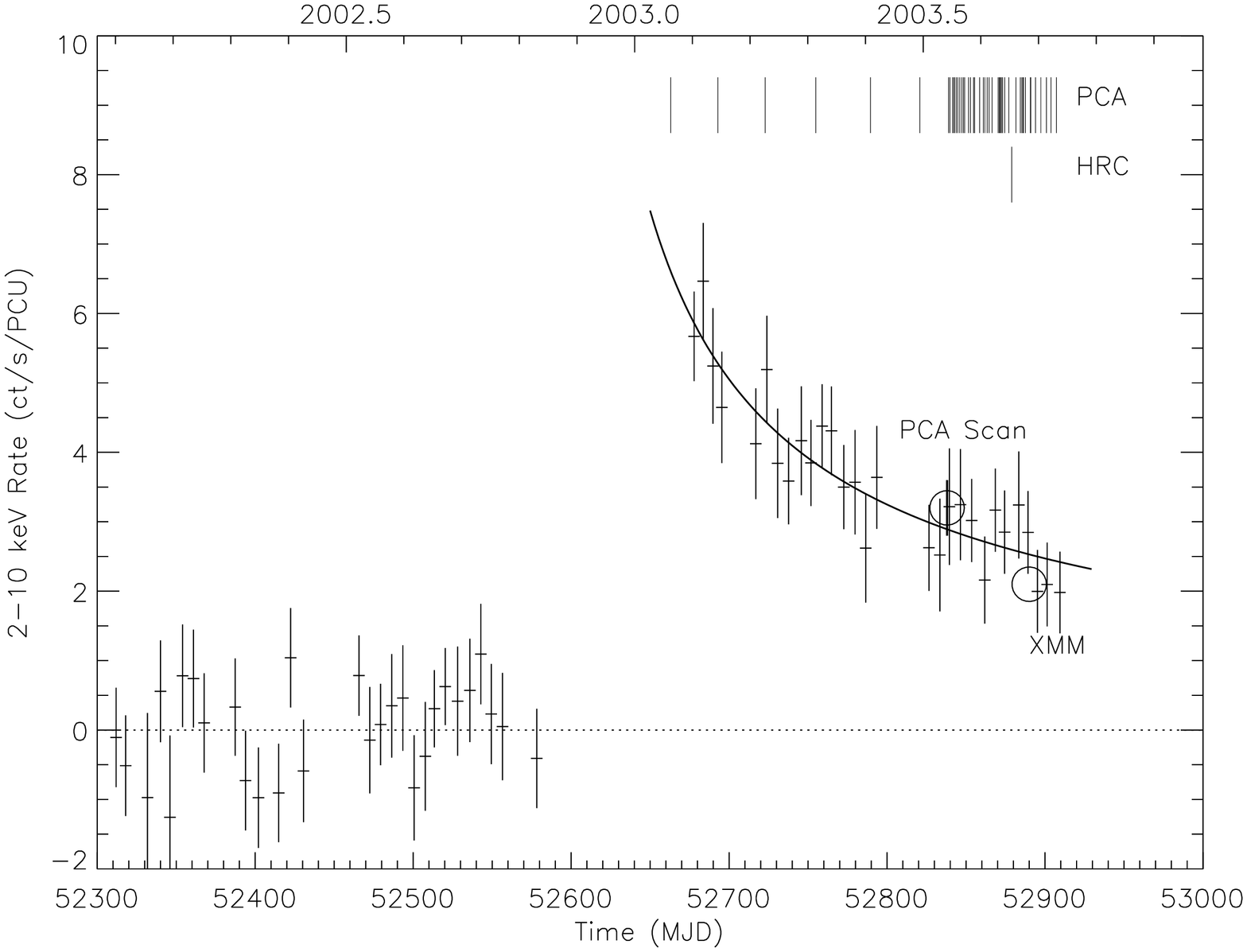}
\figurenum{3}
\vspace{-0.2cm}
\figcaption{Monitoring light curve of XTE~J1810--197, showing the transient outburst beginning in 2003 
(note: 1 mCrab = 2.27 ct/s/PCU = $2.4\times10^{-11}$ erg~cm$^{-2}$~s$^{-1}$; 2--10 keV). 
We have subtracted from the rate an offset of 0.68 ct s$^{-1}$ PCU$^{-1}$, which we ascribe to 
diffuse and unresolved emission in the region and not accounted for by our model.
Epochs of PCA pointed
observations with the source in the field of view are indicated in the top row of vertical bars. 
The epoch of the HRC pointing is shown separately.
The flux from the {\em XMM-Newton} spectrum, converted to an approximate PCA 
flux using the PIMMS simulator, is
shown as the lower circle.  The upper circle is the flux derived from the dedicated PCA scan.} 
\end{center}

\subsection{Spectrum}

A crude PCA spectrum was obtained by reanalyzing the light curves in
each spectral band, this time using the $Chandra$ position and allowing
a contribution from G11.2--0.3 (Markwardt, Ibrahim,\& Swank 2003).
The resulting spectrum of XTE~J1810--197 was clearly soft, 
despite large uncertainty in
the column densities for any model. For the column  fixed at $1
\times 10^{22}$ cm$^{-2}$ (typical for sources in the region and confirmed
by the {\em XMM-Newton} results), a
power-law fit has a photon index $\Gamma=4.7 \pm 0.6$, while  a black
body fit has $kT=0.94\pm0.11$ keV. The 2-10 keV absorbed flux was $5.5 \times
10^{-11}$ ergs cm$^2$ s$^{-1}$ on July 18. 
Additional PCA spectral data requiring analysis beyond the scope of this 
paper will address spectral evolution during the outburst. 

The source was observed with {\em XMM-Newton} on 2003 September 8. 
Results have been presented by Tiengo \& Mereghetti (2003)
and by Gotthelf et al. (2003b). We also 
processed the EPIC PN and MOS1 data using standard SAS
routines (the MOS2 data suffered pile-up in the full frame
mode) and obtained results consistent with theirs using the PN alone. 
A two-component power-law plus blackbody model provides a good fit 
with well constrained parameters of $\Gamma = 3.75 (3.5-4.1)$, 
$kT = 0.668 (0.657-0.678)$ keV, $n_H = 1.05 (1.0-1.13) 
\times10^{22}$ cm$^{-2}$, and $\chi^2_{\nu}=1.04$ ($\nu$=896) 
(errors are at the 3$\sigma$ level). 
The total unabsorbed flux in 0.5--8.0 keV is
1.35$\times$10$^{-10}$ erg~cm$^{-2}$~s$^{-1}$ with a 30\% contribution
from the blackbody component. 
This gives a source luminosity of $1.6\times10^{36} d_{10}^{2}$
erg s$^{-1}$, with $d_{10}$ the distance in units of 10 kpc. 

We analyzed archival $ROSAT$ and $ASCA$ data of the source and obtained results 
consistent with those reported by Gotthelf et al. (2003b). We find the source in a faint 
quiescent state with a flux two orders of magnitude lower than that first detected in early 
2003 and a spectrum much softer than seen above ($kT\approx0.15$~keV).

\subsection{Timing: Frequency History and Spin-down Rate}

Timing analysis was performed using a variety of PCA observations that
included pointed observations dedicated to XTE~J1810--197 (Obsid 80150-06)
observations of G11.2-0.3 and PSR J1811--1925, SGR 1806--20 (Obsids
80149-02, 80150-01), plus the bulge scans (Obsids 80106, 70138).  The
total exposure time was about 216 ks between 2003 January 23 and
September 25.  Folded light curves were extracted (2--7 keV; top PCU
layers) based on a trial folding period.  A sinusoidal profile fit
well, and was used to estimate the pulse times of arrival (TOAs) and
uncertainties.  By using a combination of all data sets we were able
to extend a phase connected solution through the complete time span.
While we attempted several models, a polynomial was conceptually
simplest.
\begin{center}
\includegraphics[scale=0.8,angle=0]{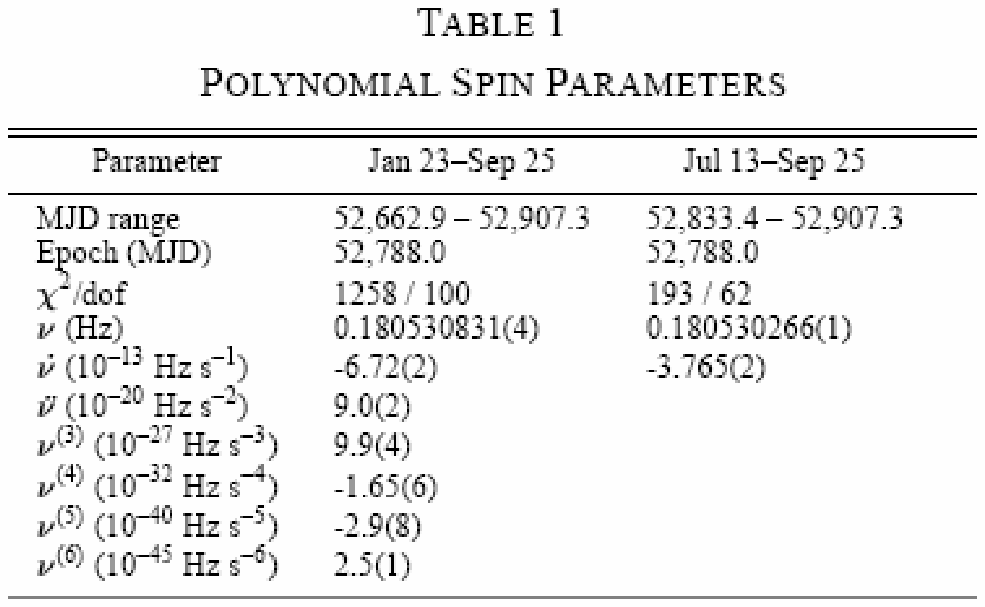}
{\footnotesize \\  Note-- Errors were determined with $\chi^2$ normalized to dof.}
\end{center}
\begin{center}
\vspace{-0.4cm}
\includegraphics[scale=0.34,angle=90]{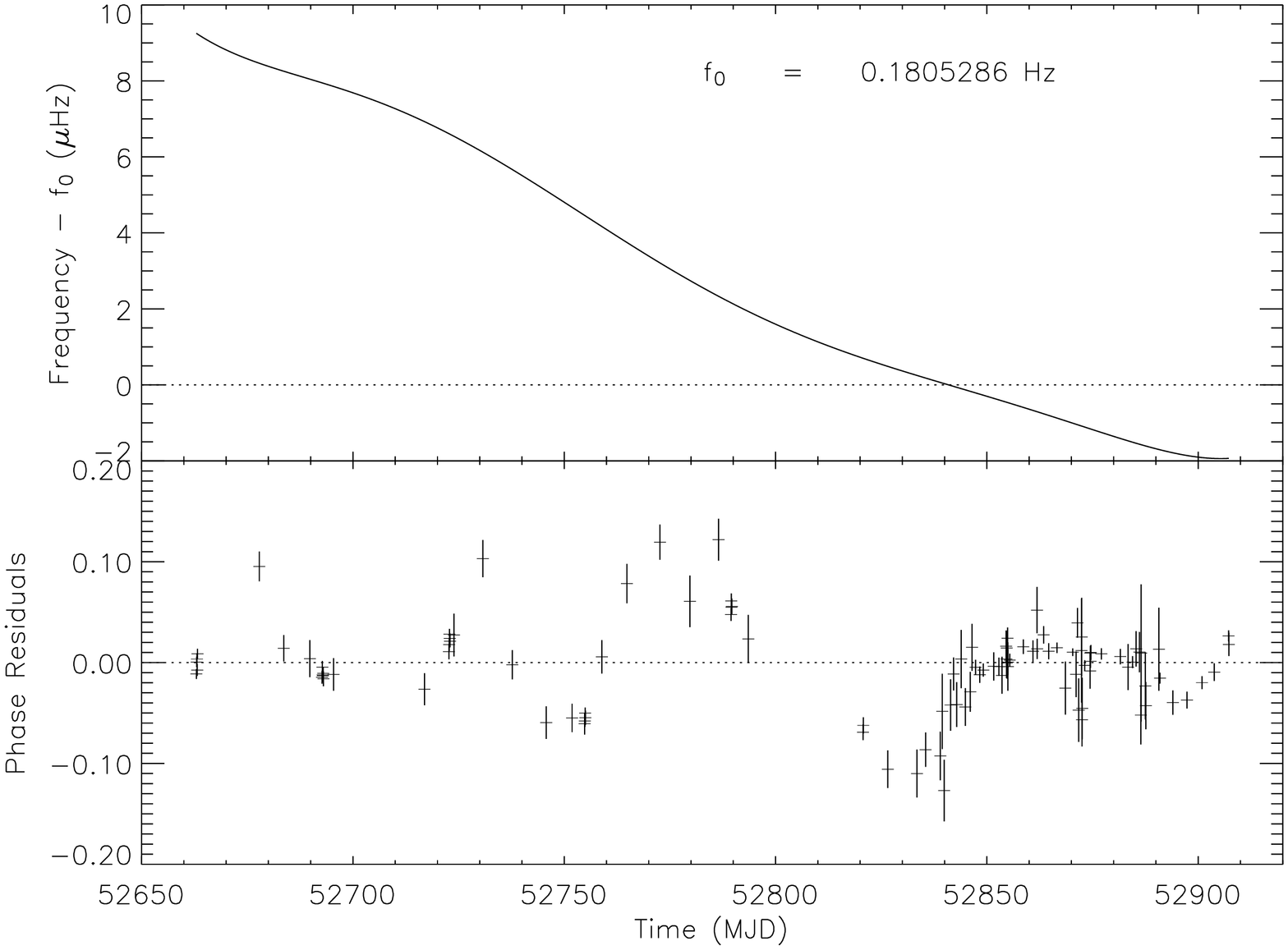}
\vspace{-0.3cm}
\figurenum{4}
\figcaption{(top) Frequency evolution and (bottom) phase residuals for PCA timing solution of 
XTE~J1810--197.}
\end{center}

Fig 4. shows the frequency evolution and phase residuals for the
polynomial fit with frequency and 6 derivatives (see Table 1 for
parameters).  While the choice of polynomial order is somewhat
arbitrary, a lower order produces significantly worse residuals.  The
weighted r.m.s residuals are 165 ms.  Reminiscent of the behavior of
1E~2259+586 after a bursting episode (Kaspi et al. 2003), the spin
down is initially steeper, but evolves to a quieter and slower
spin-down. The weighted r.m.s deviation since July is only 94 ms for a
steady spin-down (i.e. 2nd order polynomial; Table 1).  The mean pulse
period derivative is $1.8\times 10^{-11}$ s s$^{-1}$ over the full
time span of the data, and $1.15\times 10^{-11}$ s s$^{-1}$ for the
July--September time span.


With 245 days of data, it is possible to rule out a long period orbit
($\geq$ 100 days) as entirely responsible for the frequency slow down
(Markwardt et al. 2003).  While a phase-connected solution is possible
for an orbit {\it plus} a spin-down, such models are utterly dominated
by the spin-down component (best fit $\dot{\nu}$ = $-5.4 \times
10^{-13}$ Hz s$^{-1}$ for an mildly eccentric orbit with a period of
232 days; compare to Table 1), and so the orbit does not appear to add
much information.

To look for short period orbits
we made Lomb-Scargle periodograms of the phase
residuals obtained from subtracting the polynomial 
model. They show no significant
peaks at the 95\% confidence level.  For orbital periods down to 20
minutes, the peak periodogram power was 21, for a maximum
orbital amplitude, $a_x \sin i$, of 
70 lt-ms. 
Such a limit is independently inferred from the high stability of the spin down rate during the past 80 days.
This
would imply a mass function of $4 \times 10^{-7} M_\odot/P_d^2$,  $P_d$
being the binary period in days.  Thus, except for orbits improbably close
to face-on, a companion mass would be restricted to be planetary in
size. 

\vspace{-0.2cm}

\section{Discussion}

The nature of a neutron star source is principally determined by the
energy mechanism that powers its emission.
The distance of XTE~J1810--197 is likely to be 5 kpc and almost certainly in the range 
3--10 kpc (Gotthelf et al. 2003b). The unabsorbed outburst 
luminosity (from \S2.4) is $(2-16)\times 10^{35}$
ergs s$^{-1}$, which is in the range of the unabsorbed luminosities 
of AXPs and SGRs. Since the outburst episode, the rotational energy loss 
due to the pulsar
spin-down, $\dot E = I \Omega \dot \Omega \approx 4\times10^{33}$ erg
s$^{-1}$ (where $I$ is the moment of inertia of a canonical neutron
star and $\Omega = 2 \pi/P$), is at least two orders of magnitude lower
than the implied X-ray luminosity. 
A binary Doppler shift can not explain the frequency trend and 
there are strong limits on the mass of any companion in a short period  orbit. 
As discussed by Gotthelf et al. (2003b), the optical and infra-red 
limits, as well as our own limit in the red,  are sufficient
to rule out interpreting the transient X-ray source 
as a distant Be-star binary. 
Furthermore, the spectrum of the source is notably softer
than the typically hard spectra of high mass
X-ray binaries.

Magnetic braking is then a candidate to dominate the spin-down.
It appears to have been variable at the start of the
outburst and to have relaxed to a relatively stable rate of $1.15 \times 10^{-11}$
s s$^{-1}$. Such a rate, for a dipole magnetic field, would imply a magnetic field 
$B=3.2\times10^{19}\sqrt{P \dot P} = 2.6 \times 10^{14}$ G and a characteristic age 
$\tau = P/2{\dot P} \leq 7600$ yr.
Such a super-critical field strength and relatively young pulsar age
are typical of magnetars. This and the close similarities between the
temporal and spectral properties of the source and those of AXPs and SGRs 
make XTE~J1810--197 a new magnetar candidate. 
The once apparent divide between AXPs and SGRs has been blurred by the
SGR-like bursts from AXPs 1E 1048.1--5937 and 1E 2259+586 (Gavriil et
al. 2002; Kaspi et al. 2003) and the AXP-like soft spectrum from SGR~
0526--66 (Kulkarni et al. 2003) and SGR~1627-41 (Kouveliotou et al. 2003).


With the exception of the transient candidate AXP AX~1845--0258,
considerable flux variability like that shown in Fig. 3 is not commonly 
observed from AXPs and SGRs in their quiescent non-bursting states.
However, both AXPs and SGRs are known to show significant
enhancement to their persistent emission flux following active
bursting episodes.  The flux may rise by more than an order of
magnitude before  relaxing back on timescales that range
from days to years.
This behavior was observed from 1E 2259+586 
(Kaspi et al. 2003; Woods et al. 2003), SGR 1900+14 
(Woods et al. 2001; Ibrahim et al. 2001; Feroci et al. 2003) and 
SGR 1627-41 (Kouveliotou et al. 2003). 
For a power-law flux decay, the index for XTE~J1810--197 falls 
within the range of those  of SGR 1627-41 (0.47) and SGR 1900+14 
(0.6-0.9). 

We searched for bursts prior to the peak activity of the source that could have been associated
with it.  We found no bursts in the PCA observations of G11.2--0.3 on 2003 January 23. 
Five SGR-like bursts were observed by experiments in IPN on 2002 December 5 and 6 (Hurley et al. 2002). 
One was well localized to SGR~1806--20 by \it Ulysses \rm and Konus-\it Wind \rm. 
The others remain unlocalized.

Alternatively, the possibility of flux variability  due
to magnetic field disturbances is also
viable in the magnetar model. Given that a magnetic field has to be
greater than $B_0\sim2\times10^{14} (\theta_{max}/10^{-3})^{1/2}$~G to
fracture the crust and cause burst activity (Thompson \& Duncan
1995; $\theta_{max}$ is the crust yield strain), the energy associated
with fields $B<B_0$ may excite magnetospheric currents or dissipate in
the crust (Thompson, Lyutikov \& Kulkarni 2003), causing a
long-lasting flux enhancement. 

If the magnetar birth rate is  $\approx$ 10 \% of neutron stars
(Kulkarni \& Frail 1993; Kouveliotou et al. 1994),
there should be $\approx$ 100 in the same part of the galaxy to which we have yet been 
sensitive, rather than the 9 sources currently known. 
The difference may be explained if other magnetars 
are usually in a quiescent state like that of XTE~J1810--197 before the current outburst. 

\vspace{-0.4cm}

\acknowledgements
We thank C. Thompson for useful discussions and S. Snowden and M. Still for help with XMM data. AII is supported by 
NASA grants NAG5-13186 \&  NAG5-13740; the McGill groups by NSERC Discovery and Steacie grants, Canada Research 
Chairs, NATEQ, CIAR \& NASA; SS-H by NSERC UFA and Discovery grants \& NASA LTSA NCG5-356; PMW by NASA 
grant NAG5-11608; and KH by \it Ulysses \rm grant under JPL Contract 958056.

\end{document}